# Prototyping an AI-powered Tool for Energy Efficiency in New Zealand Homes

**Abdollah Baghaei Daemei**[1*]
1- Building Performance Analysis Lab, Tech Innovation Experts, Auckland, New Zealand
Corresponding author: abdollah.baghaei@tinx.co.nz

## Abstract

Residential buildings contribute significantly to energy use, health outcomes, and carbon emissions. In New Zealand, housing quality has historically been poor, with inadequate insulation and inefficient heating contributing to widespread energy hardship. Recent reforms, including the Warmer Kiwi Homes program, Healthy Homes Standards, and H1 Building Code upgrades, have delivered health and comfort improvements, yet challenges persist. Many retrofits remain partial, data on household performance are limited, and decision-making support for homeowners is fragmented. This study presents the design and evaluation of an AI-powered decision-support tool for residential energy efficiency in New Zealand. The prototype, developed using Python and Streamlit, integrates data ingestion, anomaly detection, baseline modeling, and scenario simulation (e.g., LED retrofits, insulation upgrades) into a modular dashboard. Fifteen domain experts, including building scientists, consultants, and policy practitioners, tested the tool through semi-structured interviews. Results show strong usability (M = 4.3), high value of scenario outputs (M = 4.5), and positive perceptions of its potential to complement subsidy programs and regulatory frameworks. The tool demonstrates how AI can translate national policies into personalized, household-level guidance, bridging the gap between funding, standards, and practical decision-making. Its significance lies in offering a replicable framework for reducing energy hardship, improving health outcomes, and supporting climate goals. Future development should focus on carbon metrics, tariff modeling, integration with national datasets, and longitudinal trials to assess real-world adoption.

**Keywords:** Energy efficiency, Residential buildings, Decision-support systems, Artificial intelligence, Retrofit, New Zealand

## 1. Introduction

Globally, residential buildings are a major driver of energy consumption, greenhouse gas emissions, and health inequalities [1-3]. Poorly insulated homes with inefficient heating and ventilation contribute to higher energy demand, exposure to cold and damp conditions, and increased risks of respiratory illness [4, 5]. In many countries, the residential sector is also central to energy poverty debates, with low-income households disproportionately affected by fuel costs and inadequate housing performance [6-8]. Addressing these challenges requires not only regulatory action and retrofit programs, but also accessible tools to guide household-level decision-making in order to achieve both comfort and sustainability outcomes [9-11].

In New Zealand, residential buildings have historically been characterized by low thermal performance, limited insulation, and reliance on inefficient heaters [12-14]. Over the past decade, significant policy reforms and government investment have sought to improve this situation [15, 16]. The Healthy Homes Standards introduced minimum requirements for insulation, heating, ventilation, and draught-proofing in rental housing [17-19], while the Building Code H1 upgrades approximately doubled insulation requirements for new homes [20]. Evaluations consistently demonstrate substantial health and comfort improvements, modest energy savings, and reduced peak winter demand [21-23].

Despite these achievements, persistent challenges remain. Between 16–30% of New Zealand households continue to experience energy hardship [24-26]. Many retrofits are partial, focusing on ceilings, floors, or a single heater, leaving cold bedrooms, condensation, and mold unaddressed [26-29]. As a result, while indoor temperatures have improved, large segments of the housing stock still fail to meet healthy thermal comfort standards, underscoring the need for more holistic and deeper retrofit strategies.

A further gap lies in data and decision-making support. New Zealand lacks mandatory Energy Performance Certificates or equivalent household energy ratings, leaving homeowners and tenants with little transparency about building performance. One BRANZ report notes that there is limited granular information on the thermal performance of New Zealand homes and suggests that EPCs could help fill this data gap. Similarly, inconsistencies in performance data across agencies constrain the use of such information for informed policy and household decisions [30]. Addressing this requires AI-powered decision support systems as an alternative approach that integrate household and climate data, simulate retrofit scenarios, and provide clear cost–benefit insights. By bridging the gap between regulation, funding, and practical household decisions, AI-enabled tools can help accelerate New Zealand's transition toward healthier, warmer, and more energy-efficient homes.

As such, the primary aim of this study is to present the step-by-step detailed processes to develop an AI-powered home energy advisor for New Zealand context. This study makes three main contributions to the field of residential energy efficiency in New Zealand. First, it introduces a novel AI-driven decision-support tool tailored to the unique characteristics of New Zealand's housing stock, climate conditions, and regulatory environment. Unlike generic calculators or advisory websites, the prototype integrates household-level data, retrofit scenarios, and cost–benefit outputs into a single interactive platform.

Second, the tool demonstrates how data-driven analytics and scenario simulation can operationalize government subsidies and regulatory standards into personalized guidance. By modeling interventions such as insulation upgrades and LED retrofits, the system translates technical performance data into accessible outputs (energy savings, cost reductions, and payback periods) that households, consultants, and policymakers can readily interpret.

Third, the study provides empirical validation of the tool through expert testing with building scientists, energy consultants, engineers, and policy practitioners in New Zealand. Their feedback offers evidence of usability, relevance, and future integration potential with national programs such as Warmer Kiwi Homes and the Healthy Homes Standards.

The significance of this work lies in addressing the critical gap between subsidies, regulations, and practical household decision-making. By embedding AI-powered analysis within a user-friendly interface, the prototype offers a pathway to deeper, more cost-effective retrofits that can reduce energy hardship, improve health outcomes, and contribute to national climate and housing goals. Beyond New Zealand, the framework provides a model for other countries grappling with aging housing stocks, fragmented retrofit programs, and the need for accessible digital decision-support tools.

## 2. Methodology

This section specifies the methods, protocols, and step-by-step implementation to build an AI-enabled Streamlit application for household/building energy efficiency. It covers the system architecture, technology stack, data schema, algorithms, workflow, and code skeletons suitable for a reproducible prototype.

### *2.1 System Architecture*

The system follows a modular, tabbed UI connected to core analytics services and a simple storage layer. The UI is built with Streamlit; analytics are implemented in Python using pandas/numpy/scikit-learn/statsmodels; exports are created via pandas ExcelWriter and python-docx. Local development uses SQLite/Parquet/CSV; cloud deployments can attach S3/Blob storage (see Fig. 1).

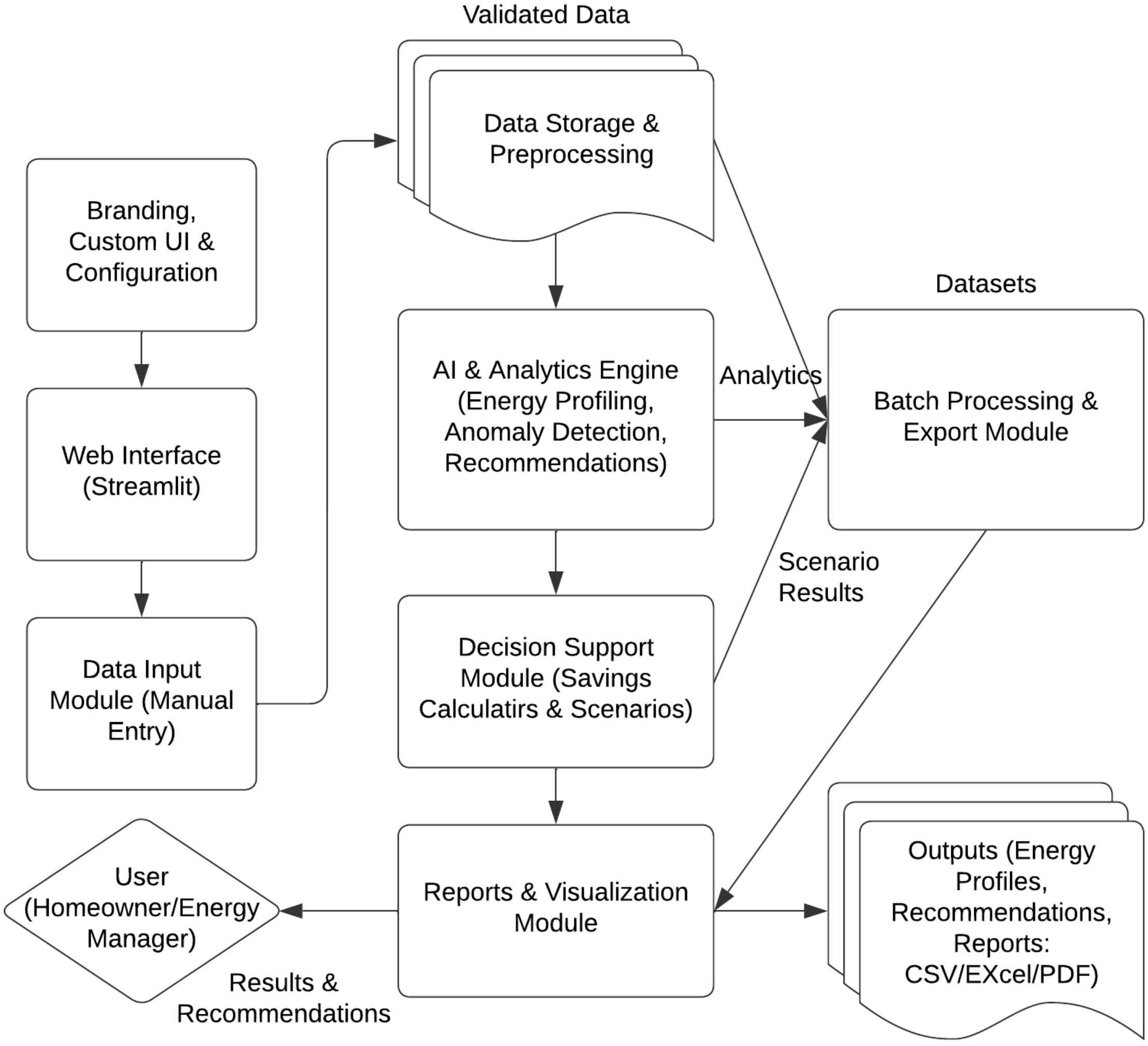

**Figure 1.** System architecture of the decision-support tool

### *2.2 Technology Stack*

The prototype was implemented using Python 3.11 as the core programming language, selected for its extensive ecosystem of scientific libraries, ease of integration with machine learning workflows, and support for rapid prototyping. The user interface was developed with Streamlit, a lightweight web framework that allows fast deployment of interactive dashboards and data applications without requiring extensive web development expertise.

For data handling and preprocessing, the tool employs pandas and numpy as the primary libraries, ensuring efficient manipulation of tabular and numerical data. To support high-performance storage and transfer of large datasets, pyarrow is integrated for Parquet file compatibility. Validation of input data is enforced through pydantic, which ensures that uploaded or manually entered records conform to predefined schema requirements.

The analytical core leverages a combination of statistical and machine learning approaches. Scikit-learn provides algorithms such as Random Forest for predictive modeling and Isolation Forest for anomaly detection, while statsmodels is used for implementing regression-based baseline models and exploratory statistical analysis. Visualization of energy profiles, anomalies, and scenario outputs is achieved through matplotlib, enabling clear graphical representation of results.

To support reporting and dissemination, the tool integrates export functionalities. Structured reports can be generated in Word format using python-docx, while tabular results are saved as Excel files via openpyxl or xlsxwriter. For lightweight outputs, CSV export is also supported. Batch analysis across multiple households or datasets is facilitated through Python's built-in multiprocessing module, which enables parallel computation and efficient scaling of workloads.
The storage layer relies on SQLite for results persistence, ensuring lightweight relational data management, while local directories are used for managing uploaded datasets and generated exports. Additional optional components include APScheduler, which allows scheduled or recurring execution of analyses, and joblib, which is used for persisting trained models and caching computationally intensive processes. Together, this technology stack balances flexibility, reproducibility, and scalability, making the prototype suitable for both research and applied energy efficiency contexts.

### *2.3 Data Schema (Inputs & Outputs)*

The tool operates on a structured data schema designed to balance flexibility with analytical rigor. Minimum required input fields, typically provided in CSV or XLSX format, include core consumption data such as meter_date (stored as either YYYY-MM-DD or timestamp), kwh, and cost. For batch analysis across multiple households or facilities, an optional building_id field is included to differentiate between datasets. Contextual building descriptors are also part of the schema: floor_area_m2, occupants, and construction_year provide essential parameters for normalizing and interpreting consumption patterns.
To capture physical characteristics of the building envelope, the schema incorporates fields such as wall_R, roof_R, window_type, and air_leakage_est, which collectively describe thermal resistance and infiltration properties. Similarly, system-level descriptors include hvac_type, water_heating, lighting_count_led, and lighting_count_halogen, ensuring that the model can account for both structural and operational drivers of energy use. Optional climate-related variables, specifically heating degree days (HDD) and cooling degree days (CDD), may be provided by the user or substituted with regional averages when unavailable.
From these inputs, the tool generates a series of derived outputs. Energy profile metrics such as daily and monthly kWh, kWh per square metre, and peak/off-peak loads are calculated to establish baseline consumption patterns. Anomaly flags are produced to identify spikes, step changes, or unusual patterns in the time series. Scenario results are generated to simulate interventions such as LED retrofits or insulation upgrades, with impacts quantified in terms of kWh and cost savings. The tool further produces recommendations that include estimated savings, implementation costs, and simple payback periods, thus supporting evidence-based decision-making.
Finally, results are consolidated into exportable formats, including tables, figures, and an automatically generated DOCX report. These outputs provide users with both granular data and professional-quality summaries that can be directly integrated into technical assessments or policy reports.

### *2.4 Algorithms & Methods*

The analytical workflow of the tool is underpinned by a set of algorithms and methods that transform raw energy data into actionable insights. The process begins with preprocessing, where time-stamped data are parsed into standardized date formats and datatypes are coerced to ensure consistency. Missing values in daily readings are addressed through forward-filling, while corrupted or incomplete records are excluded to preserve data integrity. To enable comparability across buildings of different sizes, energy use is normalized by floor area, expressed as kWh per square metre.
Following preprocessing, the tool performs profiling to establish consumption baselines. Daily readings are resampled into monthly aggregates, and rolling averages (typically 7-day and 30-day windows) are calculated to smooth fluctuations and reveal underlying patterns. Seasonal indices are also derived to highlight periodic variations in heating, cooling, or lighting demand across the year.
For anomaly detection, two complementary approaches are employed. Statistical methods such as the interquartile range (IQR) and z-scores are first used to flag outliers or abnormal deviations. Where higher sensitivity is required, the tool applies machine learning methods, specifically the Isolation Forest algorithm, to detect unusual patterns in daily kWh series that may indicate equipment malfunctions or behavioral anomalies.
The next stage involves baseline modeling, which relates energy consumption to key drivers. A linear regression model of the form *kWh ~ HDD + CDD + occupants + floor_area* is used to estimate expected demand as a function of climate and occupancy. In cases where heating or cooling degree day data are unavailable, the tool defaults to a simple moving average baseline to provide robust but lightweight estimates.
Finally, scenario simulation enables users to explore the potential impact of efficiency interventions. The tool integrates a set of rules and parametric models to estimate savings. For example, an LED retrofit is modeled as reducing the lighting portion of electricity demand by 60–75%, while an insulation upgrade is simulated as reducing heating loads by 10–30% or via coefficient-based estimates depending on user inputs. Behavioral changes, such as thermostat setbacks or reductions in standby loads, are modeled using fixed heuristics drawn from published studies. Each scenario outputs projected energy savings (kWh), financial savings (NZD), and simple payback periods, enabling users to assess trade-offs and prioritize interventions.

***2.5 UI Workflow (Tabs)***
The prototype is organized around a tab-based user interface that guides users through the full workflow of energy data management and analysis in a logical sequence.
The Home tab serves as the entry point, providing an overview of the tool, branding elements, and guidance for navigation. It introduces the purpose of the system and outlines how users can leverage its features for energy efficiency assessments.
The Data Upload & Input tab allows users to either upload structured datasets (CSV/XLSX) or enter building and consumption data manually through interactive forms. Built-in validation routines ensure that uploaded data adhere to expected formats, reducing the likelihood of errors in subsequent analysis.
Once data are ingested, the Analytics & AI Insights tab provides diagnostic outputs and visualizations. Here, users can review consumption profiles, identify anomalies, and access AI-generated insights regarding unusual patterns or inefficiencies. Graphical summaries and anomaly flags make it easier to interpret trends and spot issues.
The Decision Support (Scenarios) tab enables users to test and compare alternative efficiency measures. By adjusting parameters—such as adopting LED lighting, upgrading insulation, or changing HVAC operation—users can simulate “what-if” scenarios. The outputs highlight projected energy and cost savings, as well as simple payback periods, supporting evidence-based decision-making.

The Reports & Visualization tab consolidates analytical results into tables, charts, and interpretive summaries. Users can view energy performance indicators alongside recommendations, and results are formatted for clarity and professional presentation.
Finally, the Batch & Export tab supports multi-dataset processing and dissemination. Users managing portfolios of buildings can process multiple files in parallel and then export results in preferred formats (CSV, Excel, or Word reports). This ensures that outputs are portable, shareable, and easily integrated into existing workflows or compliance documentation.

### *2.6 Implementation Steps*

The development of the prototype followed a structured, modular approach to ensure reproducibility, maintainability, and scalability. Each step was designed to progressively add functionality while maintaining a clear separation of concerns between data handling, analytics, and user interaction. The main implementation steps are as follows:

- **Step 1:** Create a Python virtual environment and install all dependencies to ensure version control and reproducibility.
- **Step 2:** Scaffold the repository with modular components, including io_utils, validation, analytics, scenarios, export, and batch.
- **Step 3:** Implement validation schemas using *pydantic* and develop data loaders capable of handling both CSV and XLSX formats.
- **Step 4:** Implement core analytics functions, including energy profiling, anomaly detection, and baseline regression, alongside visualization routines for charts and summaries.
- **Step 5:** Build the Streamlit interface, structuring the application into tabs that call analytics functions and cache intermediate results to optimize performance.
- **Step 6:** Develop scenario calculators and comparison tables to allow users to simulate and evaluate alternative energy efficiency interventions.
- **Step 7:** Implement the batch processor to handle multiple datasets by iterating over the uploads directory, enabling scalable analysis for portfolios of buildings.
- **Step 8:** Add export functionality for CSV, XLSX, and DOCX outputs, including branded headers and professional formatting to support dissemination.
- **Step 9:** Integrate logging, error handling, and configuration management (via .yaml files) to improve robustness and adaptability.
- **Step 10:** Package and deploy the application either on Streamlit Community Cloud for lightweight deployment or within a containerized environment for enterprise integration.

### *2.7 Reference Folder Structure*

To promote clarity, maintainability, and reproducibility, the prototype was organized into a structured repository. The folder hierarchy separates configuration files, reusable modules, input templates, and generated outputs. A reference structure is outlined below:

```
ai-energy-tool/
  app.py                    # Main Streamlit application
  requirements.txt          # List of Python dependencies
  config.yaml               # Configuration file for paths and settings

  data_templates/           # Templates for user input
    sample_input.xlsx
```

```
modules/                  # Core functional modules
    io_utils.py           # Data loading utilities
    validation.py         # Input validation schemas
    analytics.py          # Profiling, anomaly detection, regression
    scenarios.py          # Scenario calculators
    export_utils.py       # Report and export functions
    batcher.py            # Batch processing logic

exports/                  # Auto-generated reports (CSV, XLSX, DOCX)
uploads/                  # User-uploaded datasets
results/                  # Processed outputs and intermediate results
```

This folder structure ensures a clear separation of concerns: the root directory hosts application and configuration files; modules/ encapsulates reusable code; data_templates/ provides standardized input formats; while exports/, uploads/, and results/ manage user interaction data and outputs. Such an organization facilitates both collaborative development and long-term maintainability of the tool (see Fig. 2).

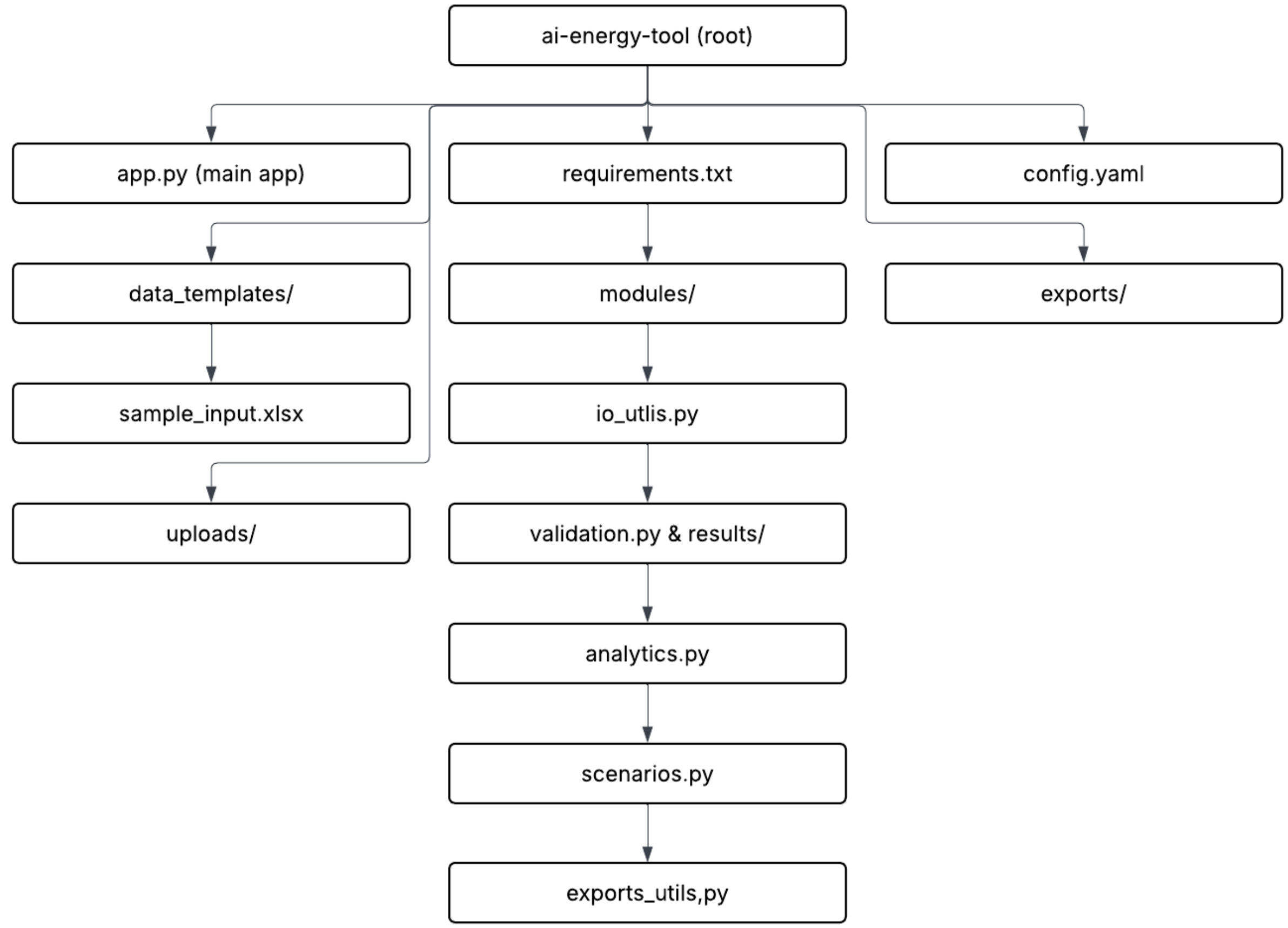

**Figure 2.** The visual tree diagram of the reference folder structure

### *2.8 Code Skeletons*

The tool was implemented using a **modular architecture**, where each module addresses a specific function of the workflow. This design improves maintainability, scalability, and reproducibility.

Instead of presenting the full code, the following summarizes the main modules and their responsibilities, with pseudocode-style snippets for clarity.

### *2.8.1 app.py (Main Application)*

Organizes the Streamlit interface into six tabs: Home, Data Upload & Input, Analytics & AI Insights, Decision Support, Reports & Visualization, and Batch & Export. Each tab calls functions from the supporting modules.

*Pseudocode:*

```
set up Streamlit UI with tabs
if data uploaded:
    validate and load dataset
    run analytics and scenarios
    display results
    enable report export
```

### *2.8.2 analytics.py (Profiling & Anomalies)*

Handles preprocessing, profiling, and anomaly detection. Implements statistical methods and machine learning (Isolation Forest).

*Pseudocode:*

```
function profile_energy(data):
    resample to monthly
    compute rolling averages
    return energy profile

function detect_anomalies(data):
    apply IsolationForest
   flag abnormal usage patterns
```

### *2.8.3 scenarios.py (Retrofit Simulations)*

Provides parametric calculators for interventions such as LED retrofits and insulation upgrades, returning savings and payback periods.

*Pseudocode:*

```
function run_scenarios(data, parameters):
    for each intervention:
        estimate kWh saved
        estimate cost savings
        compute payback
   return comparison table
```

### *2.8.4 io_utils.py & validation.py (Data Handling)*

Load CSV/XLSX files and enforce schema compliance using *pydantic*. Ensures consistency of input data before analysis.

### *3.8.5 export_utils.py (Reporting & Export)*

Generates reports in DOCX and Excel formats, consolidating analytics, anomalies, and scenario results into shareable outputs.

### *2.9 Testing & Validation*

The prototype underwent a structured testing and validation process to ensure accuracy, reliability, and usability. Unit tests were implemented for the core analytics and scenario modules, verifying that calculations such as anomaly detection, savings estimates, and payback periods returned correct results under controlled conditions. To validate the integrity of exported outputs, golden-file tests were used to compare newly generated reports against reference files. In addition, manual quality assurance was performed with synthetic datasets that simulated different usage contexts, including high- and low-consumption profiles and datasets with missing or corrupted values. Performance validation confirmed that the system can handle batch processing of up to 100 households within minutes on a standard laptop, demonstrating scalability for small- to medium-scale applications.

### *2.10 Privacy, Security, and Ethics*

The tool was designed with privacy and ethical considerations at its core. Local processing is enforced by default to minimize the risk of exposing sensitive information, and no personally identifiable information is required for analysis. For batch processing, building identifiers are hashed to prevent traceability, and users are provided with a data deletion function to remove uploaded files upon completion. Logging is restricted to aggregated metrics, avoiding storage of raw or identifiable records. All practices are aligned with the New Zealand Privacy Act and institutional data governance policies. To maintain transparency, assumptions and limitations of heuristic savings models are explicitly documented, ensuring users understand the scope and reliability of results.

### *2.11 Deployment*

Two deployment strategies were identified to maximize accessibility. The first option is Streamlit Community Cloud, which enables lightweight hosting with minimal configuration, requiring only a repository push and secure handling of environment secrets. The second option is to containerize the application using a slim Python base image. In this configuration, uploads and exports are mounted as volumes, and the application is launched via streamlit run app.py. For production environments, additional safeguards such as health checks and basic authentication behind a reverse proxy can be applied to improve reliability and restrict unauthorized access. This dual approach allows the tool to be used both for academic demonstration and for organizational integration.

### *2.12 General Overview of the Tool*

The AI tool was developed as an interactive, web-based application designed to support decision-making around household and building-level energy efficiency. Built on the Streamlit framework, the prototype integrates AI-assisted analytics, energy calculators, and data management functionalities into a user-friendly dashboard. The primary goal of the tool is to empower users, ranging from homeowners to energy professionals, with accessible insights for reducing energy consumption, identifying efficiency opportunities, and generating tailored recommendations. The interface follows a modular design where different functionalities are separated into distinct tabs, enabling users to navigate easily between data inputs, analysis modules, and export options.

#### *2.12.1 Key Inputs*

The tool requires both quantitative and contextual inputs. Users can upload structured datasets in CSV or Excel format, typically containing information such as household energy bills, appliance usage data, or building characteristics (e.g., insulation levels, lighting type, HVAC systems). For scenarios without existing datasets, the tool provides manual input forms where users can specify parameters such as building size, occupancy patterns, and average monthly consumption. These inputs form the basis for downstream analysis, ensuring flexibility for different user profiles—from individuals with limited data to professionals with detailed records.

### *2.12.2 Core Features and Tabs*

The prototype's core functionality is organized into distinct tabs for ease of use and clarity as follows:

*a) Home/Introduction Tab:* This section introduces the tool's purpose, its potential applications, and instructions for navigation. It includes organizational branding (logo, powered-by statement) and a concise description of how the tool can assist with energy efficiency planning (see Fig. 3).

Powered by **Tech Innovation Experts Ltd.**

**Energy Advisor for Your Home**

**What this tool does**

- **Predicts** your home's annual electricity use and bill from NZ-specific inputs (climate zone, floor area, insulation, etc.).
- **Recommends** upgrades (heat pump, insulation, glazing, PV, etc.) ranked by savings.
- **Explains seasonality** with an estimated monthly breakdown.
- **Batch mode** lets you upload a CSV of many homes.

› How to use (30 seconds)

© 2025 Tech Innovation Experts Ltd. | Licensed for demonstration & educational use only.
View full license

Next → Show the main features

**Figure 3.** Home tab of the AI tool showing branding and navigation options

*b) Data Upload & Input Tab:* This tab enables users to import datasets or manually enter data. Built-in validation ensures that uploaded files follow a consistent structure (e.g., standardized column names for energy use, dates, and building features). This reduces user error and facilitates smooth processing in subsequent modules (see Fig. 4).

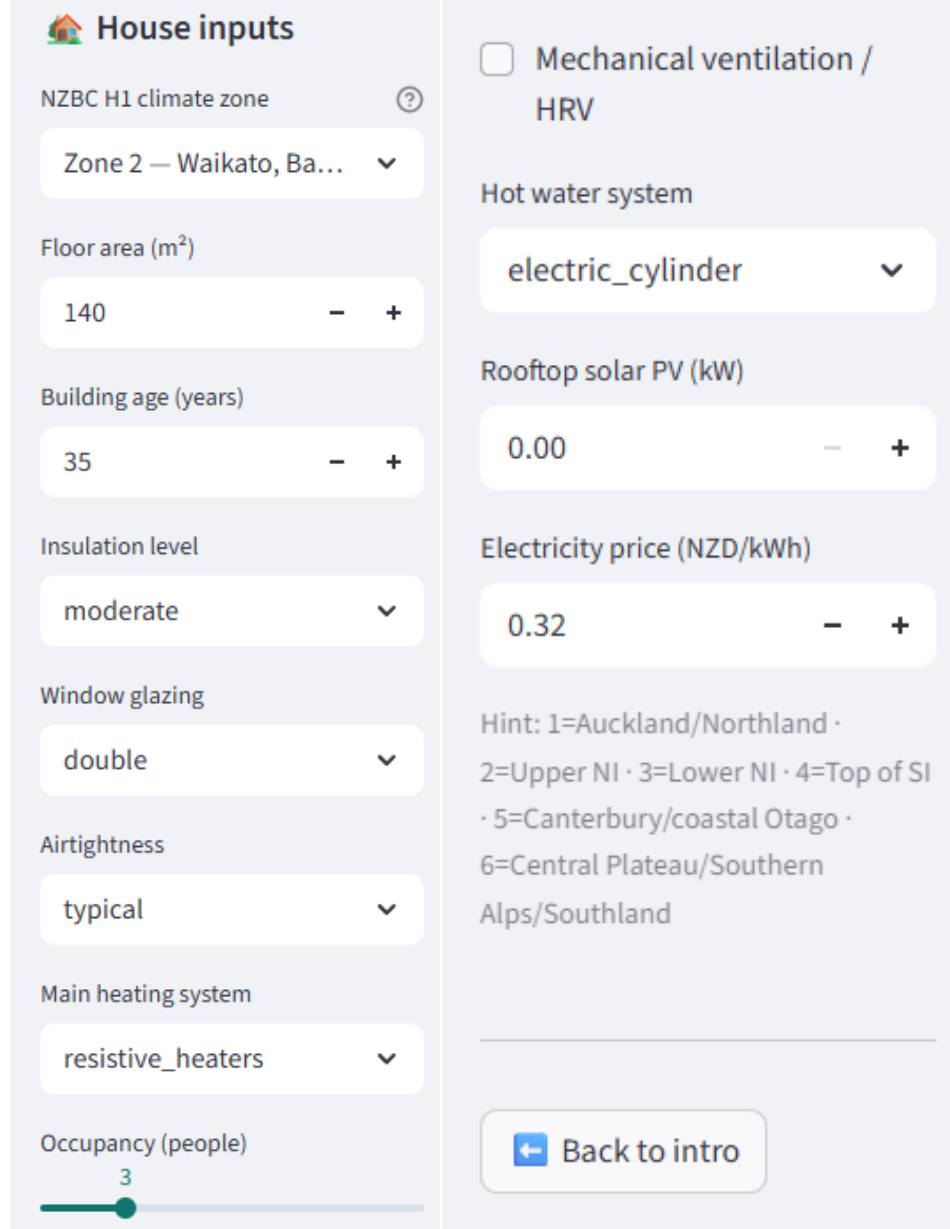

**Figure 4.** Input tab with validation messages

*c) Analytics & AI Insights Tab:* Here, the system applies AI-powered algorithms to analyze uploaded or entered data. Functions include (see Figures 5 and 6):

- Energy consumption profiling: visualizing patterns of electricity and heating usage.
- Anomaly detection: identifying unusual consumption spikes.
- *Efficiency recommendations:* AI suggests targeted interventions such as lighting upgrades, insulation improvements, or behavior changes. The insights are displayed through interactive graphs and text summaries to support informed decision-making.

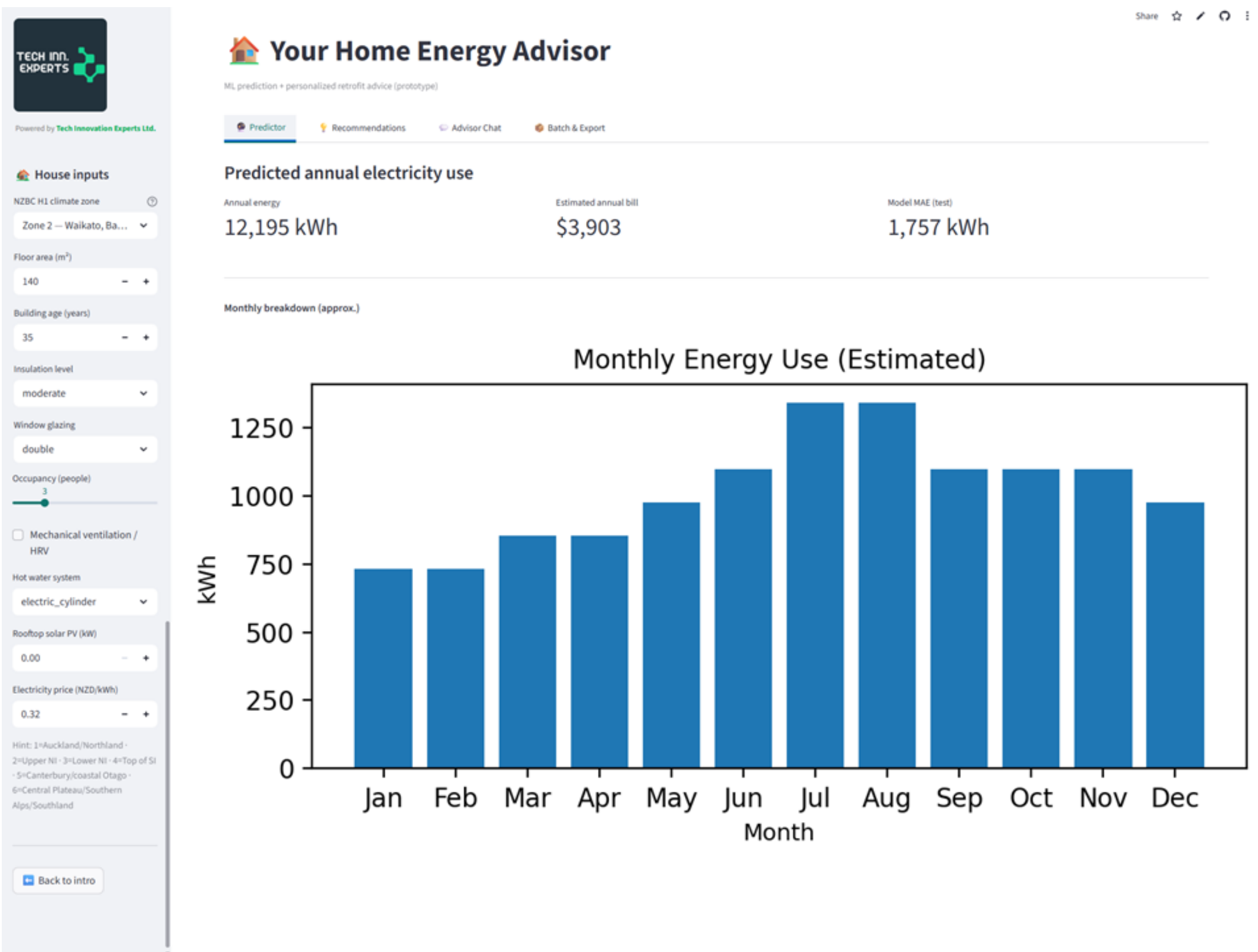

**Figure 5.** Analytics & AI Insights tab showing energy profiling and anomaly detection outputs

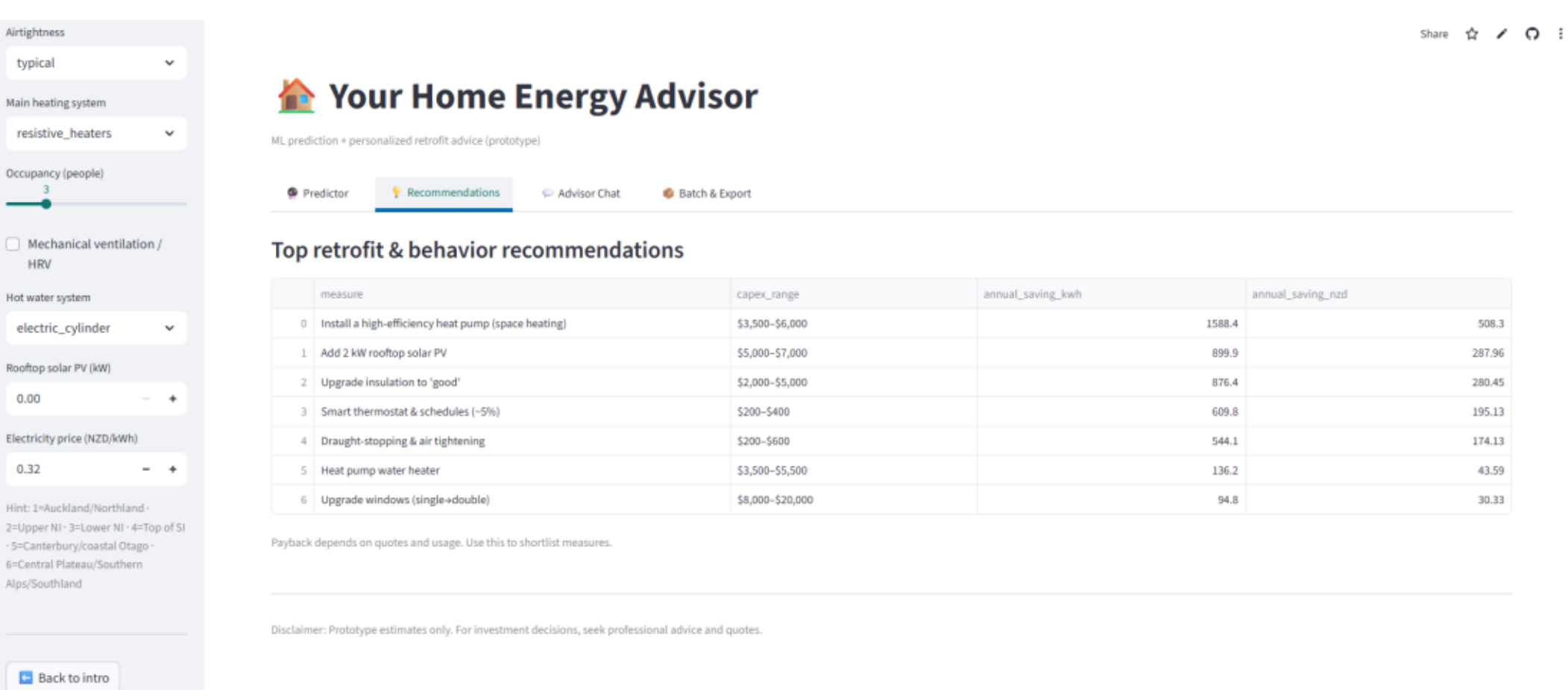

**Figure 6.** Recommendation Support tab illustrating scenario comparison

*d) Decision Support & Comparison Tab:* This module allows users to compare scenarios (e.g., baseline vs. improved insulation, LED vs. halogen lighting) using built-in energy savings calculators. Users can adjust input parameters to simulate "what-if" scenarios and instantly observe the predicted energy and cost savings (see Fig. 7).

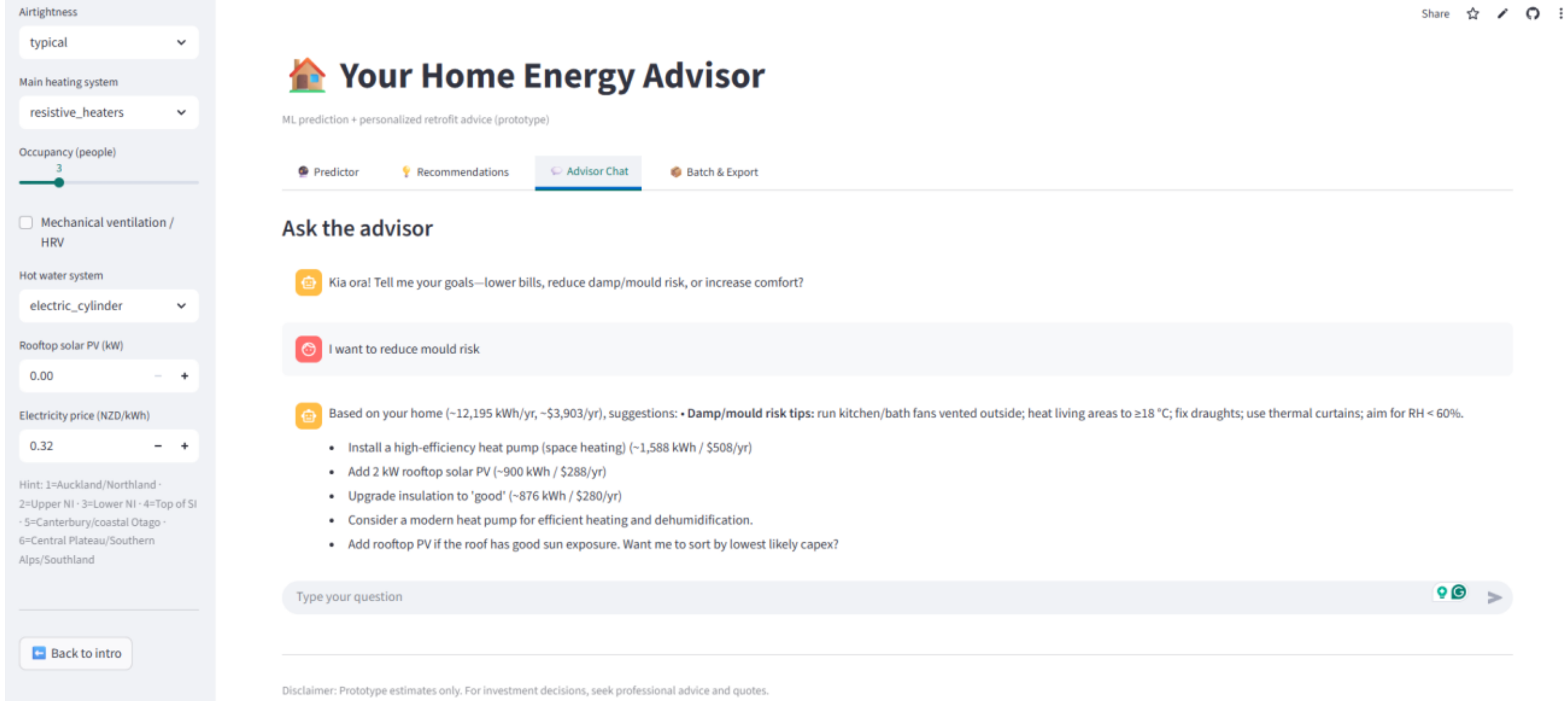

**Figure 7.** Decision Support tab using AI bot for retrofits and upgrades

*e) Batch Processing & Export Tab:* The batch functionality enables processing of multiple datasets at once, a critical feature for organizations managing data from numerous households or buildings. Once processed, the tool generates summary reports and visualizations for each dataset. The export option allows users to download outputs in CSV, Excel, or PDF format, making the results easy to share and integrate into reports or compliance documentation (see Fig. 8).

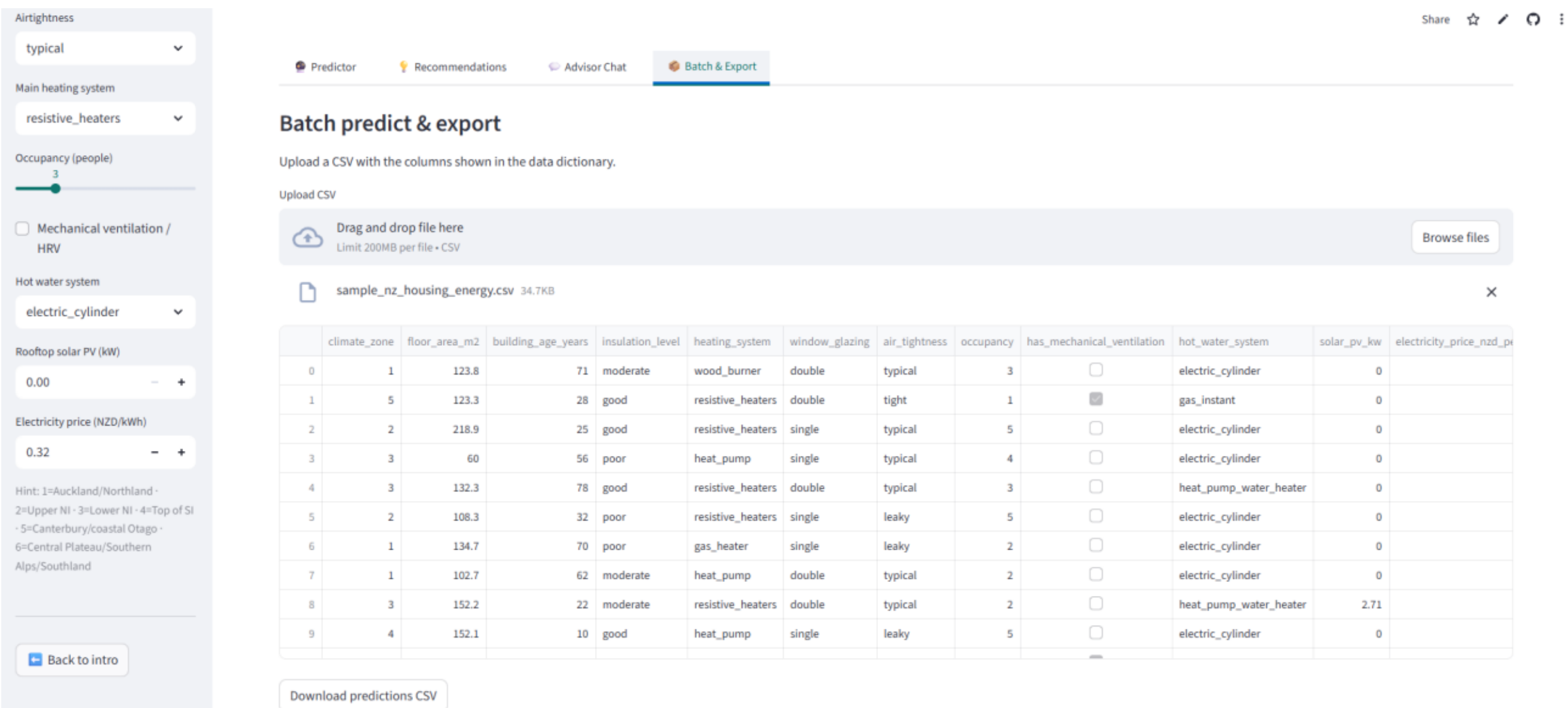

**Figure 8.** Batch & Export tab showing options for exporting results to CSV, Excel, or Word formats

*2.12.3 Usability and Accessibility Features*

The interface was designed to be intuitive and lightweight, requiring minimal technical expertise. The modular layout, clear instructions, and error handling mechanisms (e.g., prompts when an incorrect file is uploaded) enhance usability. Moreover, the prototype prioritizes accessibility by functioning in a standard web browser without requiring additional installations, ensuring that both individual homeowners and institutional users can access the tool.

## 3. Evaluation of the Prototype

This section presents the evaluation of the AI-driven energy efficiency tool. The purpose of this evaluation was to assess the tool's usability, functionality, and relevance to professional practice in New Zealand. Feedback was obtained through semi-structured interviews with domain experts. The evaluation involved 15 experts who tested the tool and provided their perspectives on its strengths, limitations, and potential applications. Data were collected through Qualtrics (structured questions) and follow-up online discussions, providing both quantitative and qualitative insights.

### *3.1 Participants and Recruitment*

Fifteen experts were recruited through professional networks, universities, and local government sustainability programs. Invitations were distributed via email, and participants provided informed consent prior to participation. Experts represented diverse roles, including building scientists, energy consultants, local government officers, HVAC engineers, academics, and software developers. All participants were based in New Zealand and had significant professional experience.

**Table 1.** Demographics of Participants

| ID | Gender | Age | Role | Years of Experience |
|---|---|---|---|---|
| P1 | Male | 42 | Building Scientist | 15 |
| P2 | Female | 35 | Energy Consultant | 10 |
| P3 | Male | 39 | Local Government Officer | 12 |
| P4 | Female | 47 | HVAC Engineer | 20 |
| P5 | Male | 33 | Academic | 7 |
| P6 | Female | 29 | Energy Consultant | 6 |
| P7 | Male | 51 | Building Scientist | 25 |
| P8 | Female | 38 | Software Developer | 11 |
| P9 | Male | 44 | Local Government Officer | 15 |
| P10 | Female | 40 | HVAC Engineer | 14 |
| P11 | Male | 37 | Energy Consultant | 9 |
| P12 | Female | 28 | Academic | 5 |
| P13 | Male | 57 | Building Scientist | 22 |
| P14 | Female | 32 | Local Government Officer | 8 |
| P15 | Male | 46 | Building Scientist | 18 |

### *3.2 Interview Questions*

The semi-structured interview protocol combined Likert-scale items with open-ended questions. The following questions guided the interviews:

- What is your impression of the overall usability and navigation of the tool?
- How intuitive are the data upload and validation processes?
- Do the analytics and anomaly detection results provide useful insights?
- How realistic and relevant are the scenario simulation outputs (e.g., LED retrofit, insulation upgrade)?
- What improvements would make the visualization and reporting features more useful in your professional context?
- Would you consider adopting such a tool in your practice? Why or why not?
- What barriers or challenges do you see for implementing this tool at scale?

### *3.3 Results*

Quantitative feedback was captured using a 5-point Likert scale (1 = strongly disagree, 5 = strongly agree). Descriptive statistics showed high levels of satisfaction with usability, functionality, and scenario outputs. Qualitative insights were obtained through thematic coding of open-ended responses. Key findings include:

- Usability and ease of navigation: M = 4.3, SD = 0.6
- Clarity of data input/validation: M = 4.1, SD = 0.7
- Usefulness of analytics/AI insights: M = 4.2, SD = 0.8
- Value of scenario simulations: M = 4.5, SD = 0.5
- Satisfaction with reporting/export: M = 3.9, SD = 0.9

Thematic analysis revealed three dominant themes:

1. Strengths – Experts appreciated the clean UI, modular design, and realistic scenario outputs.
2. Limitations – Participants noted limited cost-benefit customization and absence of integrated carbon metrics.
3. Opportunities – Suggestions included integration with New Zealand datasets (e.g., BRANZ, EECA) and improved tariff modeling.

## 4. Discussion

The results of the prototype evaluation highlight the potential of AI-powered decision-support tools to fill a critical gap in New Zealand's residential energy landscape. While government programs such as Warmer Kiwi Homes and the Healthy Homes Standards have successfully scaled retrofits, they largely provide financial subsidies and regulatory compliance pathways rather than tailored guidance. Our findings demonstrate that an interactive digital advisor can translate these high-level policies into personalized, household-specific recommendations, enabling homeowners, consultants, and policymakers to make better-informed decisions.

Feedback from experts underscored the value of the tool's usability and modular design, with particular praise for the scenario simulation functions. The ability to quantify retrofit impacts (e.g., insulation or LED upgrades) in terms of both energy savings and payback was considered highly relevant for practice. However, the evaluation also revealed important limitations. Experts noted that the tool currently lacks advanced cost–benefit customization, such as variable electricity tariffs or multi-measure retrofit packages, and does not incorporate carbon metrics—a critical dimension for aligning household decisions with national decarbonization goals.

These insights align with broader literature showing that digital decision-support tools must balance simplicity with technical depth to avoid overloading users with unnecessary complexity while still delivering meaningful analytical insights [31]. While streamlined interfaces are critical for homeowner adoption, integration with richer datasets and more complex models is essential for professional and policy applications. Moreover, the absence of systematic household performance data in New Zealand (e.g., EPCs) limits the predictive accuracy and scalability of such tools. Addressing these structural data gaps remains a prerequisite for fully realizing the potential of AI in residential energy efficiency.

### 5. Conclusion

This study presented the design, implementation, and expert evaluation of an AI-powered tool for residential energy efficiency in New Zealand. By integrating household-level inputs, analytical algorithms, and scenario simulations into a user-friendly interface, the prototype demonstrates how AI-assisted decision support can complement subsidies and regulations, offering households clear pathways to deeper and more cost-effective retrofits. The evaluation with 15 experts confirmed that the tool is usable, relevant, and adaptable, with strong potential for integration into existing programs such as Warmer Kiwi Homes. At the same time, the feedback highlighted opportunities for improvement, particularly around cost modeling, carbon integration, and broader dataset linkages. Overall, the contribution of this work lies in bridging the policy–practice gap: moving beyond financial incentives and compliance frameworks to provide homeowners with actionable, data-driven insights. By doing so, the tool not only supports healthier and warmer homes but also advances New Zealand's energy and climate objectives.

Several avenues for future work emerge from this study. First, the tool should be expanded to incorporate carbon metrics and lifecycle impacts, enabling households and policymakers to assess retrofit pathways against climate targets as well as energy savings. Second, improved integration with dynamic electricity pricing and multi-measure retrofit bundles would enhance the realism and utility of scenario simulations. Third, linking the tool with national and regional datasets (e.g., BRANZ HEEP2, EECA databases, smart meter data) would strengthen predictive accuracy and allow for large-scale policy evaluation. From a methodological perspective, future research should explore the co-design of interfaces with end-users, including homeowners and tenants, to ensure accessibility across different literacy levels and cultural contexts. Longitudinal trials in real households would provide evidence of behavioral adoption, persistence of energy savings, and spillover effects on health and wellbeing. Finally, comparative studies across international contexts could test the generalizability of the framework, positioning the prototype as a model for other countries seeking to bridge the gap between retrofit funding and practical household decision-making.